\documentclass[aps,prd,a4paper,onecolumn,amsmath,showpacs,superscriptaddress,nofootinbib,preprintnumbers,notitlepage]{revtex4-1}

\usepackage{verbatim}
\usepackage[T1]{fontenc}
\usepackage[utf8]{inputenc}
\usepackage[american]{babel}
\usepackage{epsfig}
\usepackage{graphicx,subcaption,caption}
\usepackage{booktabs}
\usepackage{multirow}
\usepackage{dcolumn}
\usepackage{amsmath}
\usepackage{mathtools}
\usepackage{amsfonts}
\usepackage{amssymb}
\usepackage{epstopdf}
\usepackage{bm}
\usepackage{siunitx}
\usepackage{braket}
\usepackage{enumitem}
\usepackage{soul}
\usepackage[table]{xcolor}
\usepackage{color}
\usepackage{transparent}
\usepackage{pifont}

\definecolor{navyblue}{rgb}{0.0, 0.0, 0.5}
\definecolor{royalblue}{rgb}{0.25, 0.41, 0.88}
\definecolor{cadmiumgreen}{rgb}{0.0, 0.42, 0.24}
\definecolor{blue-violet}{rgb}{0.54, 0.17, 0.89}
\definecolor{darkviolet}{rgb}{0.58, 0.0, 0.83}
\definecolor{orange(colorwheel)}{rgb}{1.0, 0.5, 0.0}

\usepackage{hyperref}
\hypersetup{
    colorlinks=true, 
    linkcolor=royalblue, 
    citecolor=magenta}

\newcommand\be{\begin{equation}}
\newcommand\ee{\end{equation}}

\newcommand\bea{\begin{eqnarray}}
\newcommand\eea{\end{eqnarray}}

\newcommand\gev{\mathrm{GeV}}

\newcommand\mpl{M_{\rm pl}}

\newcommand{\lcdm}{\Lambda\mathrm{CDM}}

\newcommand\ie{{\it i.e.}~}
\newcommand\eg{{\it e.g.}~}

\newcommand{\alens}{A_{\rm lens}}

\usepackage{booktabs}
\usepackage{multirow}
\usepackage{dcolumn}
\usepackage{colortbl}
\newcommand\vertsp{\rule[-2mm]{1mm}{0mm} &}
\newcommand\horsp{\rule[-1.5mm]{0mm}{4.125mm}}
\newcommand\morehorsp{\rule[-2.25mm]{0mm}{6mm}}

\definecolor{magenta(process)}{rgb}{1.0, 0.0, 0.56}

\definecolor{darkspringgreen}{rgb}{0.09, 0.45, 0.27}

\definecolor{royalblue(web)}{rgb}{0.25, 0.41, 0.88}

\newcommand\aap{A\&A}                

\newcommand\cqg{CQG}									
\newcommand\jcap{J.~Cosmology Astropart. Phys.} 	

\begin{document}

\title{What is the amplitude of the Gravitational Waves background expected in the Starobinsky model ?}

\author{Fabrizio Renzi}
\email{fabrizio.renzi@roma1.infn.it}
\affiliation{Physics Department and INFN, Universit\`a di Roma ``La Sapienza'', Ple Aldo Moro 2, 00185, Rome, Italy}

\author{Mehdi Shokri}
\email{mehdi.shokri@uniroma1.it}
\affiliation{Physics Department and INFN, Universit\`a di Roma ``La Sapienza'', Ple Aldo Moro 2, 00185, Rome, Italy}

\author{Alessandro Melchiorri}
\email{alessandro.melchiorri@roma1.infn.it}
\affiliation{Physics Department and INFN, Universit\`a di Roma ``La Sapienza'', Ple Aldo Moro 2, 00185, Rome, Italy}

\date{\today}

\preprint{}
\begin{abstract}
The inflationary model proposed by Starobinsky in 1979 predicts an amplitude of the spectrum of primordial gravitational waves, parametrized by the tensor to scalar ratio, of $r=0.0037$ in case of a scalar spectral index of $n_S=0.965$. This amplitude is currently used as a target value in the design of future CMB experiments with the ultimate goal of measuring it at more than five standard deviations. Here we evaluate how stable are the predictions of the Starobinsky model on $r$ considering the experimental uncertainties on $n_S$ and the assumption of $\lcdm$.  We also consider inflationary models where the $R^2$ term in Starobinsky action is generalized to a $R^{2p}$ term with index $p$ close to unity. We found that current data place a lower limit of $r>0.0013$ at $95 \%$ C.L. for the classic Starobinsky model, and predict also a running of the scalar index different from zero at more than three standard deviation in the range $dn/dlnk=-0.0006_{-0.0001}^{+0.0002}$. A level of gravitational waves of $r\sim0.001$ is therefore possible in the Starobinsky scenario and it will not be clearly detectable by future CMB missions as LiteBIRD and CMB-S4.
When assuming a more general $R^{2p}$ inflation we found no expected lower limit on $r$, and a running consistent with zero. We found that current data are able to place a tight constraints on the index of $R^{2p}$ models at $95\%$ C.L. \ie $p= 0.99^{+0.02}_{-0.03}$.
\end{abstract}

\maketitle
\section{Introduction}
After forty years from its first appearance in the literature, the theory of primordial inflation still offers the most successful solution to some of the inconsistency of the hot big bang cosmology \ie the flatness, horizon and monopole problems \cite{Guth,Starobinsky:1980te,Sato:1980yn,Kazanas:1980tx,Brout:1977ix,Linde:1981my,Albrecht:1982wi}. It also gives a viable mechanism to seed the primordial perturbations that are needed to form the large scale structure of the Universe we see at the present time such as galaxy clusters, filaments and the anisotropies of the cosmic microwave background (CMB). 
Along with density perturbations, also tensor modes (primordial gravitational waves) are expected to be produced during inflation \cite{Lyth:1998xn,Baumann:2009ds}. The observations of such modes will not only be a smoking gun for inflation but will also confirm the quantum nature of inflationary perturbations. 
In the last decade the experimental bounds on the amplitude of primordial gravitational waves, the so-called tensor-to-scalar ratio $r$, have seen a significant improvement. An upper limit of $ r_{0.002} < 0.064 $ has recently been provided by the Planck collaboration combining Planck and Bicep2/Keck Array BK14 data \cite{Akrami:2018odb}, an order of magnitude better than the first constraints from the BICEP experiment of $r<0.72$ at $95\%$ C.L. \cite{BICEP} in 2010.
In the coming years a new generation of CMB experiments (\eg BICEP3 \cite{BICEP3}, SPT-3G \cite{SPT-3G}, CLASS\cite{CLASS} and Advanced ACTpol\cite{ACTPol}) is expected to bring the sensitivity on the amplitude of tensor modes in the range $r\sim 0.01\,\text{-}\, 0.001$. 
Traces of primordial gravity waves are also started to be sought by gravitational interferometers in search of the so-called stochastic gravitational waves background, the analogous of the CMB in terms of gravitational waves (for a recent review see \eg \cite{Caprini_2018}). Unfortunately a direct detection of the stochastic background is still missing, but an upper limit has been placed on its amplitude from the first and second observing runs of the LIGO/VIRGO collaboration \cite{LIGO_SGWB-2017,LIGO_SGWB-2019}. 
While the search for primordial gravitational waves have lead to rule out several inflationary models \cite{Planck2013_inflat,Planck2015_inflat, Akrami:2018odb}, the significant improvement in CMB probes expected in the next years could let us to better identify the physical nature of inflation. Between the inflationary models which have survived the most recent data, one of most successful (and also the first to have been conceived) is the Starobinsky $R^2$ inflation, with $R$ being the Ricci scalar, proposed by A.A. Starobinsky \cite{Starobinsky:1980te}. Interestingly the $R^2$ has also a crucial role in solving the shortcomings of $f(R)$ theories which have been proposed as one of the possible alternatives to the cosmological constant of the concordance $\lcdm$ model \cite{Motahashi:2010aa,Motohashi:2011ab,Gannouji:2009sa,Motohashi:2009ba,Tsujikawa:2009ka,Motohashi:2010rf,Motohashi:2013hga,Tsujikawa:2007dd,Appleby:2008hh,Frolov:2008tda,Kobayashi:2008ffa,Appleby:2010gtg}. 
Because of its agreement with current observations, the Starobinsky model is now considered as a "target" model for several future CMB experiments as, for example, the Simons Observatory \cite{Ade:2018sbj}, CMB-S4~\cite{CMB-S4}, and the LiteBIRD satellite experiment \cite{LBIRD}. Assuming the current best-fit values of the scalar spectral index $n_S$ from the Planck experiment, the Starobinsky model predicts a tiny tensor amplitude namely $r \simeq 0.003$ for $60$ e-folds. The goal of these future experiments is therefore to have enough experimental sensitivity to measure such signal with enough statistical significance with $\delta r < 0.001$.

However the prediction of $r \simeq 0.003$ is a first approximation that does not consider several caveats. First of all, there is an experimental uncertainty on the value of $n_S$ derived under $\Lambda$CDM and this affect the predicted value for $r$, since, for example, for higher values of $n_S$ the expected value of $r$ is smaller. Secondly, there is a severe anomaly in the Planck data on the amount of gravitational lensing present in the CMB angular spectra. The lensing signal, parameterized by the parameter $A_{lens}$, is indeed larger than what expected in the $\Lambda$CDM scenario by more than two standard deviations. Since $A_{lens}$ correlates with $n_S$, the lensing anomaly could affect the predictions on $r$. Finally, there is clearly no fundamental reason to believe that the Starobinsky model is the correct inflationary scenario and, for example, several generalization could be considered.
The goal of this paper is therefore to evaluate the amount of gravitational waves predicted by Starobinsky model considering the current uncertainties on $n_S$ and the possibility of an extension to the $\Lambda$CDM model parametrized by $A_{lens}$.

Moreover, we also consider a minimal generalization of Starobinsky inflation, the so-called $R^{2p}$ models (with $p\approx 1$). These inflationary models were first proposed by \cite{Schmidt:1989ma,Maeda:1989ff} in the context of higher derivative theories and subsequently were applied to inflation providing a straightforward and elegant generalization of the $R^2$ inflation \cite{Muller:1990ffa,Gottlober:1992lo,Deflice:2010ss,Martin:2013tda,Martin:2014uu}. While the introduction of a variable index of the Ricci scalar in the inflationary action complicates the simplicity of $R^2$ inflation it allows significant deviations from the benchmark value of the tensor amplitude of the Starobinsky model and could in principle results in a better agreement with data. 
In this paper we provide constraints on Starobinsky inflation and on the more general $R^{2p}$ model using CMB anisotropies data. In particular we make use of the publicly available Planck 2015 and Biceps2/Keck array data releases. The present work is structured as follows: 
in section I we outline the main features of the generalized Starobinsky models and we derive the expression of  the scalar spectral index, $n_S$, its running, $\alpha_S$ and the tensor-to-scalar ratio $r$ as function of the number of e-foldings, $N$ and the index $p$. In section II we describe the method employed for the comparison of the theoretical model with data, while results are reported in section III. Finally in section IV we draw our conclusions. 

\section{Theory}
\begin{figure*}[!hbtp]
	\centering
	\begin{subfigure}{.65\textwidth}
		\includegraphics[width=\textwidth,keepaspectratio]{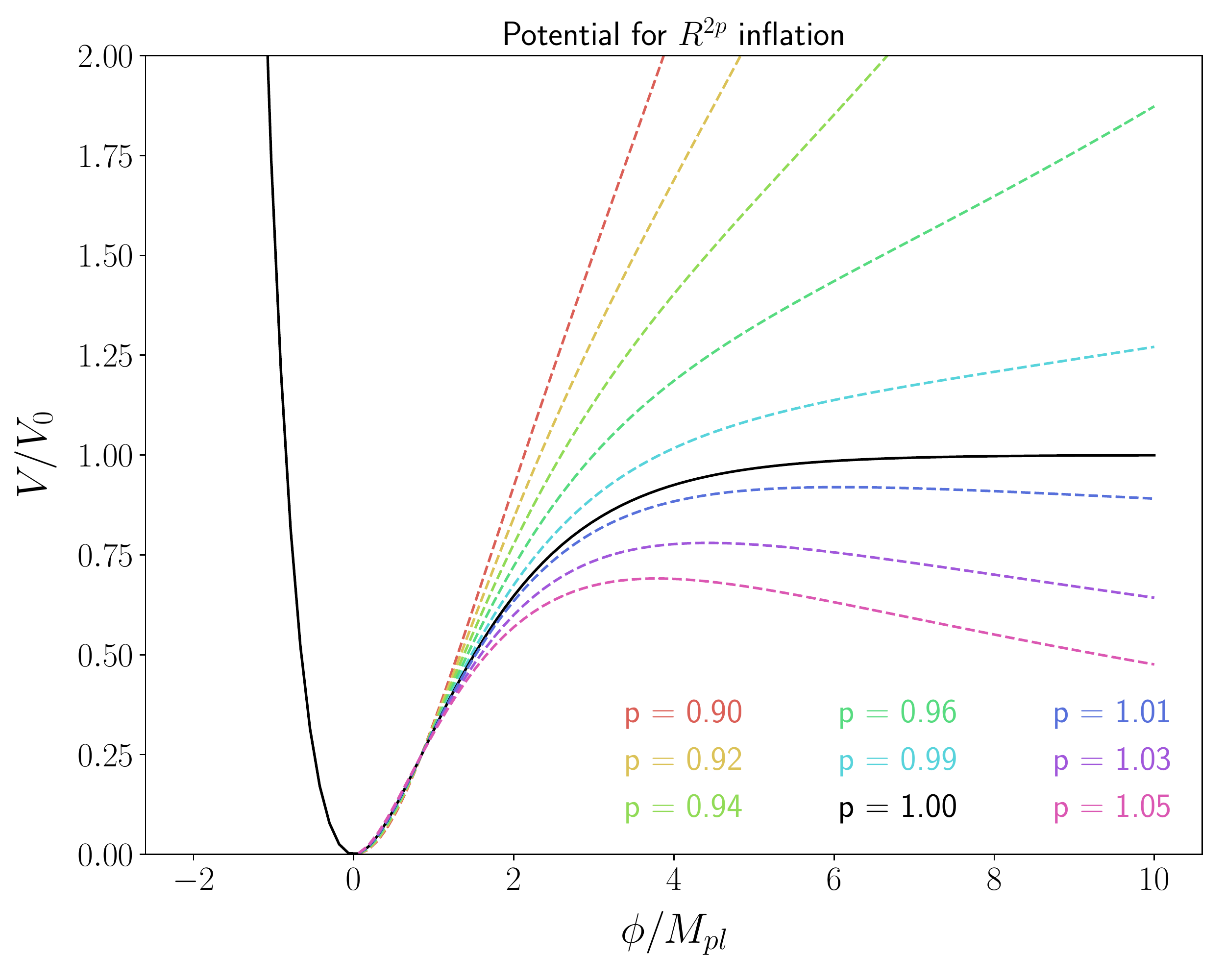}
	\end{subfigure}
	\caption{The potential of $R^{2p}$ inflation for different values of $p$.}
	\label{fig1}
\end{figure*}
\begin{figure}
    \centering
    \includegraphics[width=0.65\textwidth,keepaspectratio]{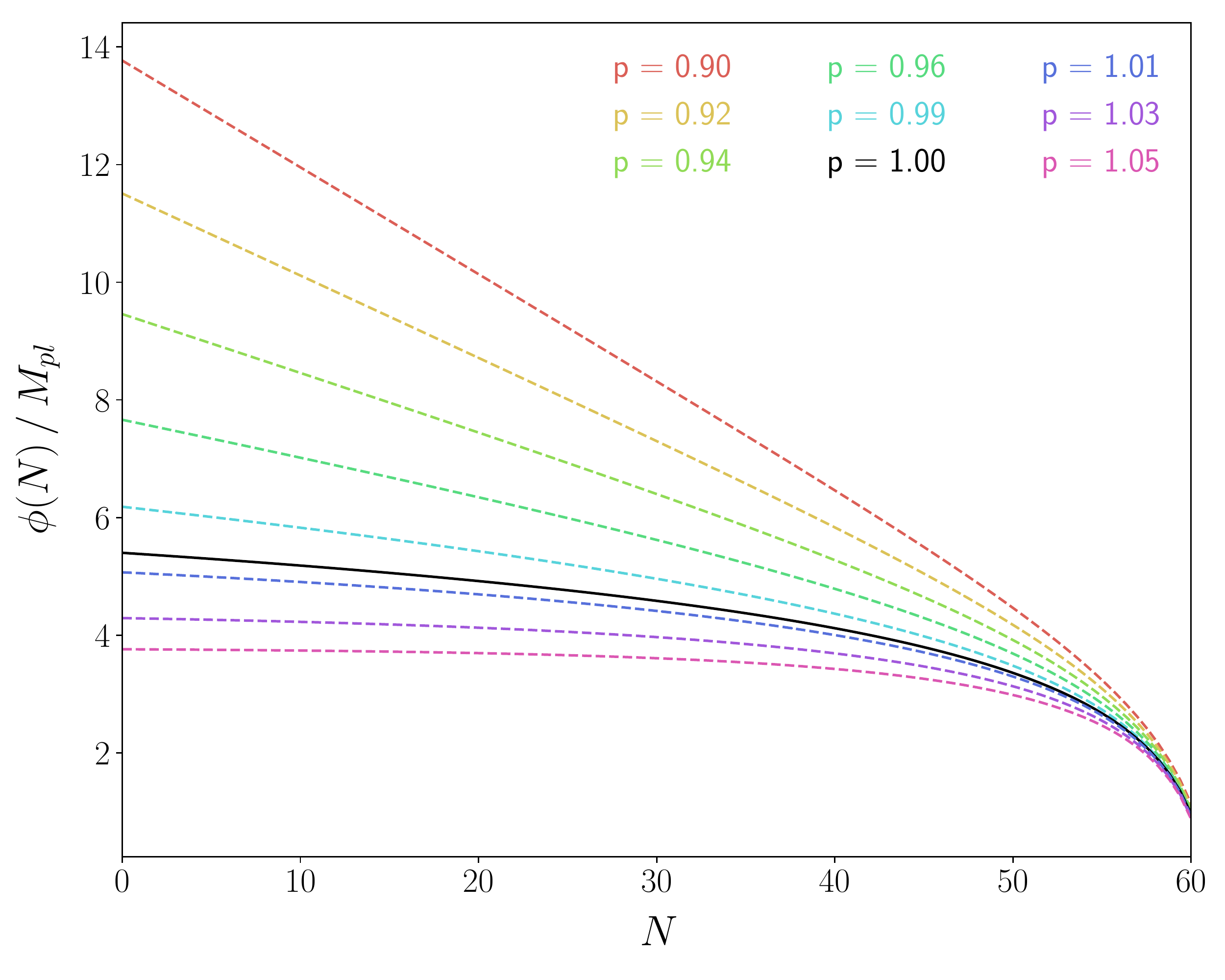}
    \caption{Time evolution of the scalar field $\phi$ for several value of the index $p$}
    \label{fig:phi_N}
\end{figure}

We start with the form of action for $R^{2p}$ inflation in the Einstein frame as:
\begin{equation}
S_{f}=-\frac{\mpl^2}{2}\int\sqrt{-g} d^{4}x f(R)
\label{action}
\end{equation}
where $\mpl = (8\pi G)^{(-1/2)}$ is the reduced Planck mass. By applying a conformal transformation of the form $g_{\mu\nu}^E = F(R)g_{\mu\nu}$ and defining a scalaron field as:
\begin{equation}\label{Eq.scalaron}
    F(R)\equiv f'(R)\equiv e^{\sqrt{\frac{2}{3}}\phi/\mpl}
\end{equation}
the above action can be rewritten in the following form \cite{Motahashi:2014okj}:
\begin{equation}
S_{\varphi}=\int d^{4}x\sqrt{-g}\bigg({-\frac{\mpl^2}{2}R_E+\frac{1}{2}g^{\mu\nu}\partial_{\mu}\phi\partial_{\nu}\phi-V(\phi)}\bigg)
\label{actione}
\end{equation}
where the potential is given by:
\begin{equation}
    V(\phi) = \frac{\mpl^2}{2}\frac{\chi F(\chi)-f(\chi)}{F(\chi)^2}
\end{equation}
where $\chi=\chi(\phi)$ is a solution of Eq.\eqref{Eq.scalaron} for $R=\chi$. In this paper we focus on model where the $f(R)$ can be written in the form:
\begin{equation}
    f(R)=R+\frac{R^{2p}}{(6M^2)^{2p-1}}
\end{equation}
where $p$ is a real number close to unity and $M \simeq 10^{13} \gev$ is a normalized energy scale from the amplitude of observed power spectrum for the primordial perturbations. In such a model the potential $V(\phi)$ assumes the form:
\begin{equation}
    V(\phi)=V_{0}e^{-2\sqrt{\frac{2}{3}}\frac{\phi}{M_{pl}}}(e^{\sqrt{\frac{2}{3}}\frac{\phi}{M_{pl}}}-1)^{\frac{2p}{2p-1}}
    \label{potential}
\end{equation}
where $V_{0}=6(\frac{2p-1}{4p})M_{pl}^{2}M^{2}(\frac{1}{2p})^{\frac{1}{2p-1}}$. 
In Fig.\ref{fig1} we report the behaviour of the potential, $V(\phi)$,  for different values of the index $p$. 
%
As shown in the figure, the behavior of the potential for the $R^{2p}$ model depends significantly on the value of the index $p$. 
\begin{itemize}
   \item For $p \lesssim  1 $ the inflationary potential has only one vacuum corresponding to the origin and there is only one regime in which inflation can proceed. 
   Therefore for $p \lesssim 1$ we have only a small deviation from a Starobinsky-like inflation with the inflationary phase ending by violation of the slow-roll conditions. 
   \item In the opposite regime, $p \gtrsim 1$, the potential has a maximum at:
    \begin{equation}\label{eq.phi_m}
        \frac{\phi_m}{\mpl} = \sqrt{\frac{3}{2}}\ln{\left(\frac{2p-1}{p-1}\right)}
    \end{equation}
and allows two different vacua, the origin and the positive infinity. However it easily shown from Eq.\eqref{Eq.scalaron} that positive infinity does lead to an unphysical inflationary regime since $R \xrightarrow{\phi \rightarrow \infty} \infty$ \cite{Liu:2018htf,Inagaki:2019hmm}. In order to avoid this regime, we should require that,
    \begin{align}
        & \phi_{i} \lesssim \phi_m \\ 
        & \dot{\phi} = \frac{d\phi}{dt} = \frac{1}{H}\frac{d\phi}{d N} < 0 
    \end{align}
in order to neglect the behavior of the inflationary potential for $\phi > \phi_m$ and discuss only the regime in which the inflaton evolves towards the true vacuum ($\phi = 0$) where, again, inflation ends by violation of the slow-roll conditions and the $R^{2p}$ model leads only to small deviations from Starobinsky inflation (we will show in the following that these conditions are satisfied for all $p > 1 $ allowing us to neglect the inflationary regime for $\phi > \phi_m$). 
\item for $p=1$ we recover the potential of Starobinsky inflation asymptotically approaching a constant value, $V_0 = 3/4 M^2\mpl^2$, for large $\phi$:
    \begin{equation}
        V(\phi) = \frac{3}{4}M^2\mpl^2\left(1 - e^{-\sqrt{\frac{2}{3}}\frac{\phi}{\mpl}}\right)^2
    \end{equation}
\end{itemize}
In the Einstein frame, the slow-roll parameters can be expressed through the potential as:
\begin{equation}
\epsilon=\frac{\mpl^2}{2}\bigg(\frac{V'(\phi)}{V(\phi)}\bigg)^{2},\qquad\quad \eta=\mpl^2\frac{V''(\phi)}{V(\phi)},\qquad\quad\zeta^{2}=\mpl^4\frac{V'(\phi)V'''(\phi)}{V^{2}(\phi)}
\label{para}
\end{equation}
where prime denotes derivative w.r.t the scalar field $\phi$. One can express the number of e-folds, between an initial time $t_i$ and $t$, as :
\begin{equation}
    N_E \equiv\sqrt{\frac{1}{2\mpl^2}}\int^{\phi_{i}}_{\phi}{\frac{V}{V'}d\phi} \equiv\sqrt{\frac{1}{2\mpl^2}} \int^{\phi_{i}}_{\phi}{\frac{1}{\sqrt{\epsilon}}d\phi}
    \label{efolds}
\end{equation}
where $\phi_i = \phi(t_i)$. It is possible to show that during the slow-roll regime the number of e-folds is approximately the same in both the Einstein and Jordan frame which allows us to drop the subscript E while we continue to follow dynamics of inflation in the Einstein frame \cite{Motahashi:2014okj}. 
Let us start by describing first the general case where $p\neq 1$. When $p\neq 1$, the slow-roll parameters of Eq.\eqref{para} for the potential of Eq.\eqref{potential} are defined as:
\begin{align}
&\epsilon=\frac{4\bigg((p-1)F-2p+1\bigg)^{2}}{3(2p-1)^{2}\bigg(F-1\bigg)^{2}}
\label{eps}\\
&\eta=\frac{4}{3(2p-1)^{2}(F-1)^{2}}[(2p^{2}-4p+2)F^2+(-10p^{2}+13p-4)F+8p^{2}-8p+2)]
\label{eta}\\
&\zeta^{2}=\frac{16}{9(2p-1)^{4}(F-1)^{4}}\bigg[\bigg((p-1)(4p^{3}-12p^{2}+12p-4)\bigg)F^4-\bigg(48p^{4}-150p^{3}+173p^{2}-87p+16\bigg)F^3 \nonumber\\
&+\bigg(148p^{4}-388p^{3}+373p^{2}-156p+24\bigg)F^2-\bigg(168p^{4}-380p^{3}+318p^{2}-117p+16\bigg)F\nonumber\\
&+4\bigg(16p^{4}-32p^{3}+24p^{2}-8p+1\bigg)
\label{zeta}
\end{align}
with $F=e^{\sqrt{\frac{2}{3}}\frac{\phi}{\mpl}}$. 
Defining the end of inflation by $\epsilon = 1$, one can obtain the value of the scalaron when inflation end:
\begin{equation}
    \frac{\phi_f}{\mpl} = \sqrt{\frac{2}{3}}\ln{\bigg[\frac{(1+\sqrt{3})(2p-1)}{4p-(1+\sqrt{3})}\bigg]} 
\end{equation}
which is a value of order unity for $p\simeq 1$. The number of e-folds between $\phi_i$ and $\phi$ can be instead derived from Eq.\eqref{efolds}:
\begin{equation}\label{eq.N_phi}
N(\phi)=-\frac{3p}{4(p-1)}\ln\bigg[\frac{(p-1)e^{\sqrt{\frac{2}{3}}\frac{\phi_{i}}{M_{pl}}}-2p+1}{(p-1)e^{\sqrt{\frac{2}{3}}\frac{\phi}{M_{pl}}}-2p+1}\bigg]
\end{equation}
We can therefore neglect the contribution of $\phi_f$ to obtain the total number of e-folds during inflation:
\begin{equation}
N_k=N(\phi_f)\simeq-\frac{3p}{4(p-1)}\ln\bigg(\frac{(p-1)e^{\sqrt{\frac{2}{3}}\frac{\phi_{i}}{M_{pl}}}}{1-2p}+1\bigg)
\label{sadeh}
\end{equation}
which can be inverted to obtain :
\begin{equation}\label{eq.in_inflaton}
\sqrt{\frac{2}{3}}\frac{\phi_{i}}{\mpl}\simeq\ln\left[\frac{(2p-1)}{(p-1)}\bigg(1-\mathcal{C}_k\bigg)\right]
\end{equation}
with $ \mathcal{C}\equiv \mathcal{C}(N,p)\equiv e^{-\frac{4N(p-1)}{3p}}$ and $\mathcal{C}_k = \mathcal{C}(N_k,p)$. Comparing  Eq.\eqref{eq.in_inflaton} with Eq.\eqref{eq.phi_m} it is straightforward to see that $\phi_i \lesssim \phi_m$ independently of the value of $p$. Finally, we can invert Eq.\eqref{eq.N_phi} to obtain :
\begin{equation}
      \frac{\phi(N)}{\mpl} = \sqrt{\frac{3}{2}}\ln{\left[e^{4 N (p-1)/3p}\left(\frac{1-2p}{p-1} + e^{\sqrt{\frac{2}{3}}\frac{\phi_i}{\mpl}} \right) - \frac{1-2p}{p-1} \right]}
\end{equation}
which we plot in Fig.(\ref{fig:phi_N}) to show that $\phi$ is always a decreasing function of time. Therefore $R^{2p}$ models allow only for small deviations w.r.t. Starobinsky inflation independently of the value assumed by the index $p$.
Armed with these relation we can eliminate the dependency from the scalaron in Eqs.(\ref{eps} -- \ref{zeta}) to obtain the slow-roll parameters as function of $p$ and $N$ only: 
\begin{align}
&\epsilon=\frac{4\mathcal{C}_k^{2}(p-1)^{2}}{3(\mathcal{C}_k(1-2p)+p)^{2}}\label{eq.epsilon_p} \\ 
&\eta=\frac{4(p-1)}{3((1-2p)\mathcal{C}_k+p)^{2}}\bigg(2\mathcal{C}_k^{2}(p-1)+p\mathcal{C}_k-p\bigg) \label{eq.eta_p} \\
&\zeta^{2}=\frac{16\mathcal{C}_k(p-1)^{2}}{9((1-2p)\mathcal{C}_k+p)^{4}}\bigg(4(p-1)^{2}\mathcal{C}_k^{3}+p(8p-7)\mathcal{C}_k^{2}-p(11p-9)\mathcal{C}_k+p(3p-2)\bigg)\label{eq.zeta2_p}
\end{align}
The final step is now to relate inflationary observables namely the scalar spectral index $n_s$, the tensor-to-scalar ratio $r$ and the running of the scalar index $\alpha_s\equiv dn_s/dln\ k$ to the slow-roll parameters. Since the inflationary observables are invariant under conformal transformations \cite{Chiba_2008,Gong_2011} we can evaluate them making use of relations we derived in the Einstein frame. Up to leading order, we can express the inflationary observables as:
\begin{equation}
n_s=1-6\epsilon+2\eta,\quad\quad\quad \alpha_{s}=16\epsilon\eta-24\epsilon^{2}-2\zeta^{2},\quad\quad\quad r=16\epsilon
\label{parameters}
\end{equation}
making use of Eqs.(\ref{eq.epsilon_p} - \ref{eq.zeta2_p}) we obtain:
\begin{subequations}\label{eq.r_ns_alphas}
   \begin{equation}
    r=\frac{64\mathcal{C}_k^{2}(p-1)^{2}}{3\left[\mathcal{C}_k(1-2p)+p\right]^{2}}   
   \end{equation}
   \begin{equation}
       n_S=1-\frac{8(p-1)\left[\mathcal{C}_k^{2}(p-1)-p(\mathcal{C}_k-1)\right]}{3\left[\mathcal{C}_k(1-2p)+p\right]^{2}}
   \end{equation}
   \begin{equation}
       \alpha_{S}=-\frac{32p\mathcal{C}_k(p-1)^{2}(\mathcal{C}_k-1)(\mathcal{C}_k-3p+2)}{9\left[\mathcal{C}_k(1-2p)+p\right]^{4}}
   \end{equation}
\end{subequations}
The consistency relations between above equations take the following form:
\begin{equation}\label{Eq.ns_r}
n_{S}-1=-\frac{(3p-2)\sqrt{r}}{\sqrt{3}p}+\frac{8(1-p)}{3p}-\frac{r(3p-1)}{8p}
\end{equation}
\begin{equation}\label{Eq.alphas_r}
\alpha_{S}=\frac{4(1-p)(3p-2)\sqrt{r}}{3\sqrt{3}p^{2}}-\frac{(15p^{2}-20p+6)r}{6p^{2}}-\frac{(3p-2)(8p-3)r^{\frac{3}{2}}}{16\sqrt{3}p^{2}}.
-\frac{(2p-1)(3p-1)r^{2}}{64p^{2}}
\end{equation} 
Now, let us consider the case $p=1$. The slow-roll parameters Eq.\eqref{para} when $p \rightarrow 1 $ reduce to
\begin{align}
&\epsilon=\frac{4}{3(e^{\sqrt{\frac{2}{3}}\frac{\phi}{M_{pl}}}-1)^{2}} \\ &\eta=-\frac{4(e^{\sqrt{\frac{2}{3}}\frac{\phi}{M_{pl}}}-2)}{3(e^{\sqrt{\frac{2}{3}}\frac{\phi}{M_{pl}}}-1)^{2}} \\
&\zeta^2=\frac{16(e^{\sqrt{\frac{2}{3}}\frac{\phi}{M_{pl}}}-4)}{9(e^{\sqrt{\frac{2}{3}}\frac{\phi}{M_{pl}}}-1)^{3}} 
\end{align}
By using the Eq. \eqref{efolds}, the total number of e-folds in this case is obtained as
\begin{equation}
N_k\simeq\frac{3}{4}e^{\sqrt{\frac{2}{3}}\frac{\phi_{i}}{M_{pl}}}
\end{equation}
Finally for the spectral index, the running spectral index and the tensor-to-scalar ratio, we have  from Eqs. \eqref{parameters}:
\begin{equation}
n_{S}=1-\frac{2}{N_k},\qquad\alpha_{S}=-\frac{2}{N_k^{2}},\qquad r=\frac{12}{N_k^{2}}
\end{equation}
\begin{figure*}[!hbtp]
	\centering
	\includegraphics[width=\textwidth,keepaspectratio]{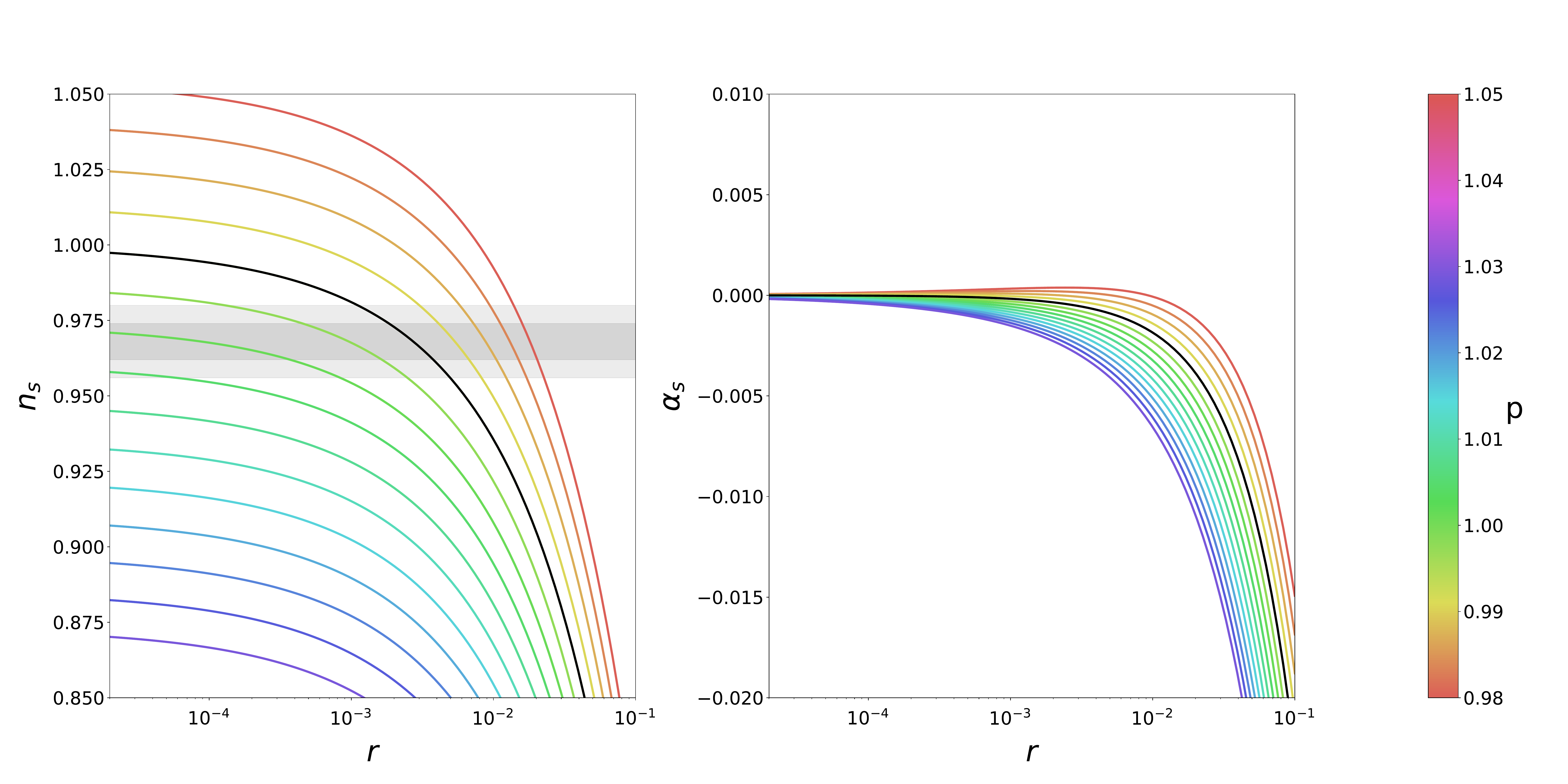}
\caption{The spectral index (left) and the running spectral index (right) versus tensor-to-scalar ratio with respect to the different values of $p$. The dashed line on the left panel shows the case for $p=1$. The gray band show the $68\% $ and $95\% $ C.L. constraints on the spectral index from Planck 2015. The black lines show the case $p=1$ }
	\label{fig2}
\end{figure*}
We show in Fig.\ref{fig2} the scalar spectral index $n_S$ (left panel) and its running $\alpha_S$ (right panel) as function of the tensor-to-scalar ratio $r$ for different value of the index $p$. We superimpose on the curves drawn according to Eqs.(\eqref{Eq.ns_r} - \eqref{Eq.alphas_r}) the Planck 2015 bounds on $n_S$ to show how the models considered in the present work can fit with observations of CMB anisotropies. We see from the left panel of Fig.\ref{fig2} that for arbitrary small values of $r$ the scalar index saturates to a maximum value which depends only on $p$, namely
\begin{equation}
n_S - 1 = \frac{8(1-p)}{3p}    
\end{equation}
for $1.01 \lesssim p \lesssim 1.02$, the saturation value falls well within the Planck bound on $n_S$, this model are therefore well in agreement with Planck data for a tensor-to-scalar ratio consistent with zero. For $p \gtrsim 1.02$ the value of $n_S$ is always outside the Planck bounds, thus we expect these models to be ruled out by current data. Model with $p \lesssim 1.01$ are within the Planck bounds only for a finite range of values of the tensor amplitude $r$ these models are not ruled out only if their range is contained in the Planck upper limit for $r< 0.1$. For $\alpha_S$ we see a similar behavior as $r\rightarrow 0$ (right panel of Fig.\ref{fig2}), but the saturation value now is zero for every value of the index $p$ since $\alpha_S \propto \sqrt{r}$ for $r\rightarrow0$. Therefore we expect that Planck data will be able to give a bound on $p$ if it is let free to vary while the bounds on $r$ and $\alpha_S$ will be consistent with zero. Conversely for the Starobinsky model we expect to have a bound on $r$ in the range $10^{-4} - 10^{-3}$ and thus an indication for a non zero running at more than two standard deviation.

\begin{table*}[!hbtp]
\begin{center}
\begin{tabular}{lcccccccc}
\toprule
\horsp
Parameter \vertsp $ \omega_b $ \vertsp $ \omega_c $ \vertsp $\theta_s$ \vertsp $ \tau$ \vertsp $ \ln(10^{10}A_s) $ \vertsp $ N $ \vertsp $ p $ \vertsp $A_{\rm lens}$\\
\hline \hline
\morehorsp
Prior   \vertsp $ [0.005 - 0.1] $ \vertsp $ [0.001-0.99] $ \vertsp $ [0.5-10] $ \vertsp $ [0.01-0.8] $ \vertsp $ [2 - 4] $ \vertsp $ [20-100] $\vertsp $ [ 0.9 - 1.05 ] $ \vertsp $ [0 - 2] $ \\
\bottomrule
\end{tabular}
\end{center}
\caption{Range of the flat prior on the parameters varied in the MCMC analysis }
\label{Tab:prior}
\end{table*}

\section{Comparison with recent experimental data and expected signal}

As stated in the previous sections the aim of this paper is to show how stable are the prediction of the Starobinsky model on inflationary parameters when a model-dependent approach is used to sample the cosmological parameter space. The general approach when looking at constraints from observations on inflationary models (see e.g. \cite{Planck2013_inflat,Planck2015_inflat,BICEP,BICEP2}) is to let the parameters $n_S$, $r$ and $\alpha_S$ free to vary assuming them to be independent from one another and then comparing the prediction of a specific model with the allowed parameter space. On the one hand, this allows to explore the inflationary sector in a model independent way but has the drawback of not allowing to sample the whole parameter space of a specific theory. Furthermore the assumption that $n_S$, $r$ and $\alpha_S$ are independent from one another is also in contrast with the prediction of any theory of inflation that assumes the validity of the slow-roll conditions (see e.g Eq.\eqref{Eq.ns_r} and Eq.\eqref{Eq.alphas_r} and also \cite{Baumann:2009ds}). 
In this work we choose a different approach: we impose an inflationary model \textit{a priori} (here, $R^{2p}$ inflation) and we extract the posterior distribution of the parameters of that specific model. 
In particular, we exploit Eqs.\eqref{parameters} to reduce the number of inflationary parameters to only two: the total number of e-folds, $N_k$, and the index, $p$.
While this approach is more model-dependant, it may results in constraints that are not achievable with the standard approach in which the inflationary parameter are independently sampled and any value of $n_S$, $r$ and $\alpha_S$ is permitted. 
The theoretical models are calculated using the latest version of the Boltzmann integrator CAMB \cite{camb}, and we use publicily available version of the Monte Carlo Markov Chain (MCMC) code CosmoMC \cite{cosmomc} (Nov 2016 version) to extract constraints on cosmological parameters. 
To compare our theoretical models with data, we use the full 2015 Planck temperature and polarization datasets which also includes multipoles $\ell < 30$. Eventually we combine the Planck likelihood with the Biceps/Keck 2015 B-mode likelihood.
We modified the code CosmoMC to include the total number of e-folds, $N_k$, and the index, $p$, as new independent parameters and to calculate the inflationary parameters $n_S$, $r$ and $\alpha_S$ throughout  Eqs.(\ref{eq.r_ns_alphas}). In what follows we will also refer to the total number of e-folds only as N dropping the subscript k.
Along with the inflationary parameters, we consider the following cosmological parameters: the baryon $\omega_b = \Omega_bh^2$ and the CDM density $ \omega_c h^2$, the angular size of the sound horizon at decoupling $\theta_S$, the optical depth $\tau$, the amplitude of scalar perturbations $A_S$ and the phenomenological lensing parameter $ A_{\rm lens} $. The flat prior imposed for these parameters are reported in Tab.\ref{Tab:prior}.

\section{Results for Starobinsky Inflation }

\begin{figure*}[!hbtp]
	\centering
	\includegraphics[width=.7\textwidth,keepaspectratio]{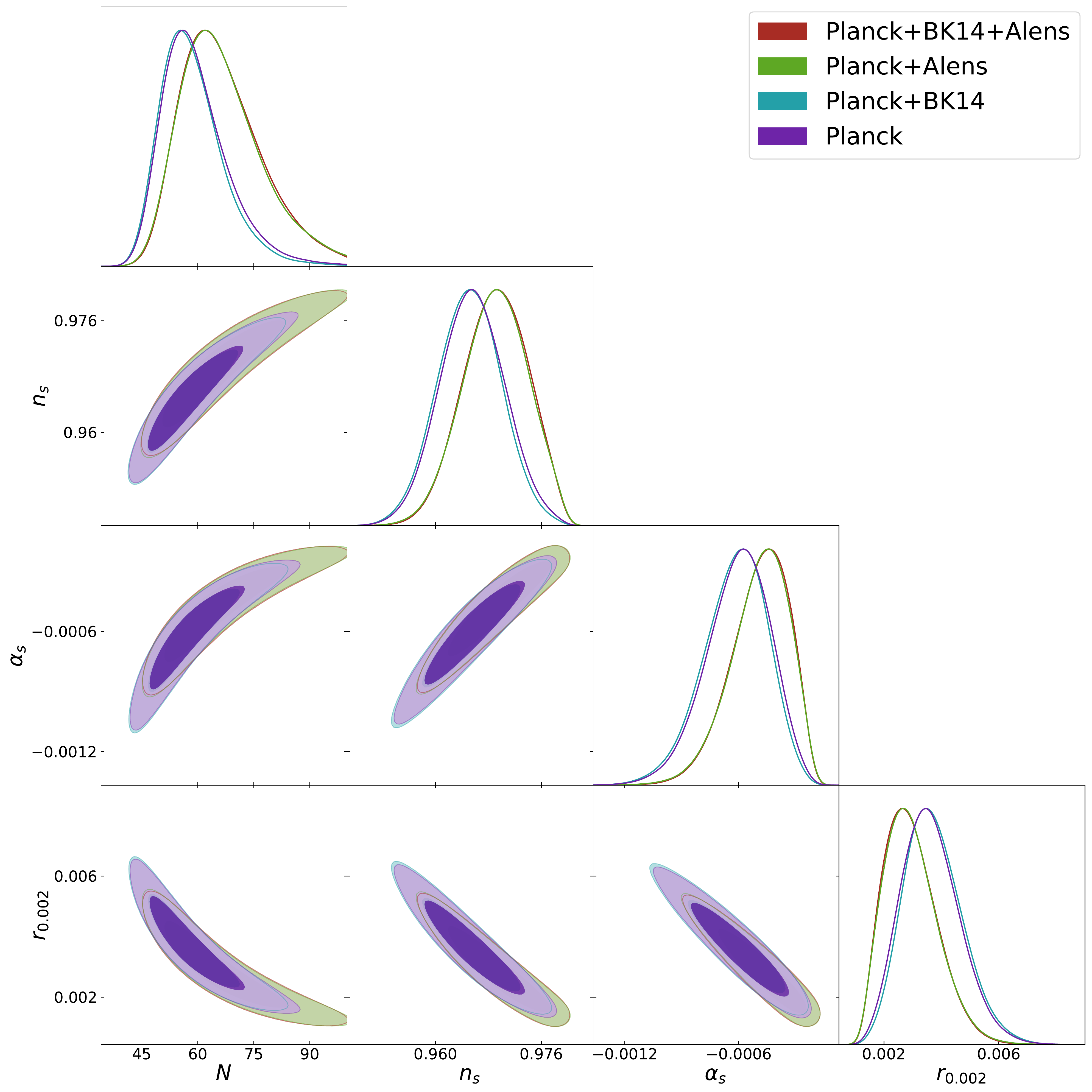}
	\caption{Constraints at $ 68\% $ and $ 95\% $ C.L. for the full Planck 2015 likelihood (Planck) and its combination with the Biceps/Keck 2015 B-mode likelihood (Planck+BK14) for the inflationary parameters for Starobinsky inflation $p = 1$.}
	\label{Fig.StarOb_results}
\end{figure*}

We report the bounds on the inflationary parameters for the Starobinsky model obtained using the full Planck 2015 likelihood (Planck) and its combination with Bicep/Keck 2015 data (Planck+BK14) in Tab.\ref{Tab.StarOb_results}. The $ 68\% $ and $ 95\% $ C.L. contour plots are showed in Fig.\ref{Fig.StarOb_results} instead. Let us start by discussing the results from the Planck datasets alone (without the inclusion of $A_{\rm lens}$). As we can see from the first column of Table \ref{Tab.StarOb_results}, we found evidence for a non-zero tensor-to-scalar ratio at the 2-$\sigma$ level when using the full Planck 2015 data ($r_{0.002} \sim 0.0036 $). This result is not coming from an actual presence of tensor perturbations in Planck data but rather it is arising from the correlation between $r_{0.002}$ and $n_S$ present in the model considered. In fact, Planck data are only able to place an upper bound on the value of the tensor-to-scalar ratio due to the poor polarization data at large scales ($r_{0.002} < 0.11$ in a one-parameter extension of the $\Lambda$CDM model) while they are able to place a strong constraint on the scalar spectral index at the accuracy of $\sim 0.6\%$ ($n_S = 0.968 \pm 0.006$) when the standard approach is used to sample these parameters. Enforcing a dependence of $n_S $ from $r_{0.002}$ therefore limits the parameter space for the tensor-to-scalar ratio and force its value to fit in the available range for $n_S$. This situation can be better understood looking at Fig.\ref{fig2}, where we show the behavior of the scalar index as a function of tensor-to-scalar ratio. The same argument can be applied to the running of the scalar index $\alpha_S$ for which we find an evidence to be negative ($\alpha_S \sim 0.0006 $) at the 3-$\sigma$ level. Again, we stress that this is not due to an indication of a running in the data but to the specific correlation which arises in Starobinsky inflation between the running and the other inflationary parameters. However these bounds show either that future measurements of $r_{0.002}$ and $\alpha_S$ have the potential to rule out the Starobinsky inflation, either that they should be considered in the analysis of future data being key parameters in studying  the feasibility of inflationary  models (see also \cite{shokri2019}). We can see from Fig.\ref{Fig.StarOb_results} and the third column of Table \ref{Tab.StarOb_results} that the combination of BK14 and Planck data do not significantly modify the bounds coming from the Planck datasets alone. The main reason for this is that the combination of Planck and Biceps2 data is compatible with every value of the tensor-to-scalar ratio satisfying $r_{0.002} < 0.07$ \cite{Planck2015_inflat} and therefore is not able to improve the constraints of Planck data alone since the bounds on $r_{0.002}$ now fall well within this limit. It is worth noting that, the slight decrease in the best-fit value of $ r_{0.002} $ when including BK14 is caused by an increase in the best-fit value of the reionization optical depth that requires a smaller scalar spectral index which in turns demand a smaller tensor ratio and a more negative running. We see from Fig.\ref{Fig.StarOb_results} and the second column of Table \ref{Tab.StarOb_results} the addition of the parameter $A_{\rm lens}$ leads to changes in the best-fit of all other parameters while not affecting their bounds. Here, the main difference with our base model is an increasing in $n_S$ of the $0.4\%$ and a reduction of $1.8\%$ of the scalar amplitude $A_S$. This in turn leads to a reduction of the optical depth $\tau$ from $0.08$ to $0.06$. To account for this shift, Planck data requires $A_{\rm lens} > 1 $ to give more smoothing on the acoustic peaks of the scalar spectrum than in the base $\Lambda$CDM model (see e.g. \cite{Planck2013_params,Planck2015_params} for a more detailed discussion). The parameters $N$, $\alpha_S$ and $r_{0.002}$ best-fit values are consequently shifted due to the correlation with $n_S$ introduced by Starobinsky inflation. The combination of Planck and BK14 data do not significantly modify the situation described here, since again the bound on $r_{0.002}$ are around an order of magnitude smaller than the sensibility of the two datasets $\delta r \sim 10^{-1}$. 
It is worth noting that both for Planck alone and for Planck+BK14 the inclusion of $A_{lens}$ provides a better fit to the data with $\Delta\chi^2 = 4$ again underlying the preference for more lensing power in Planck data.

\begin{table*}[!hbtp]
	\centering
	\begin{tabular}{lcccc}
		\toprule
		\horsp
		    & Planck \vertsp Planck+$A_{\rm lens}$ \vertsp Planck+BK14 \vertsp Planck+BK14+$A_{\rm lens}$ \\
		\hline\hline
		\morehorsp
		$\Omega_{\mathrm{b}} h^2$               \vertsp $ 0.02226\pm 0.00016 $ 				  \vertsp $ 0.02241\pm 0.00017 $               \vertsp $ 0.02224\pm 0.00016 $            \vertsp $ 0.02242\pm 0.00017$         \\  
		$\Omega_{\mathrm{c}} h^2$               \vertsp $ 0.1196\pm 0.0015 $ 				  \vertsp $ 0.1182\pm 0.0016 $                 \vertsp $ 0.1198\pm 0.0015 $              \vertsp $ 0.1182\pm 0.0015 $         \\  
		$\ln(10^{10} A_\mathrm{s}) $            \vertsp $ 3.096\pm 0.033 $ 				      \vertsp $ 3.045\pm 0.041 $                   \vertsp $ 3.104\pm 0.032 $                \vertsp $ 3.045\pm 0.040 $         \\  
		$ N $  			                        \vertsp $ 59^{+6}_{-10} $ 				      \vertsp $ 66^{+8}_{-10} $                    \vertsp $ 59^{+6}_{-10} $                 \vertsp $ 67^{+8}_{-10} $         \\  
		$ n_S $  		                        \vertsp $ 0.9656\pm 0.0048 $ 			      \vertsp $ 0.9691^{+0.0053}_{-0.0047} $       \vertsp $ 0.9652\pm 0.0048 $              \vertsp $ 0.9692\pm 0.0049 $       \\
		$\alpha_S $                         \vertsp $ -0.00060^{+0.00019}_{-0.00014} $    \vertsp $ -0.00049^{+0.00018}_{-0.00012} $   \vertsp $ -0.00062^{+0.00019}_{-0.00015}$ \vertsp $ -0.00049^{+0.00018}_{-0.00012} $\\
		$ r_{0.002} $                           \vertsp $ 0.00363^{+0.00085}_{-0.0011} $	  \vertsp $ 0.00294^{+0.00070}_{-0.0011} $     \vertsp $ 0.00371^{+0.00089}_{-0.0011} $  \vertsp $ 0.00292^{+0.00070}_{-0.0011} $  \\
		$ \tau $  		                        \vertsp $ 0.081\pm 0.017 $ 			          \vertsp $0.057\pm 0.020 $                    \vertsp $ 0.084\pm 0.017 $                \vertsp $ 0.057\pm 0.020 $       \\	
		$ \chi^2 $                              \vertsp $ 12948 $ 						      \vertsp $ 12944 $                            \vertsp $ 13594 $                      \vertsp $ 13590 $     \\
		\bottomrule
	\end{tabular}
    \caption{Constraints on inflationary parameters for a Starobinsky inflation ($ p = 1 $) from the Planck and Planck+BK14 datasets with and without the inclusion of the parameter $ A_{\rm lens} $. Constraints on parameters are at the $ 68\% $ C.L.}
    \label{Tab.StarOb_results}
\end{table*}

\section{Results for near-Starobinsky inflation }
\begin{figure*}[!hbtp]
	\centering
	\includegraphics[width=.75\textwidth,keepaspectratio]{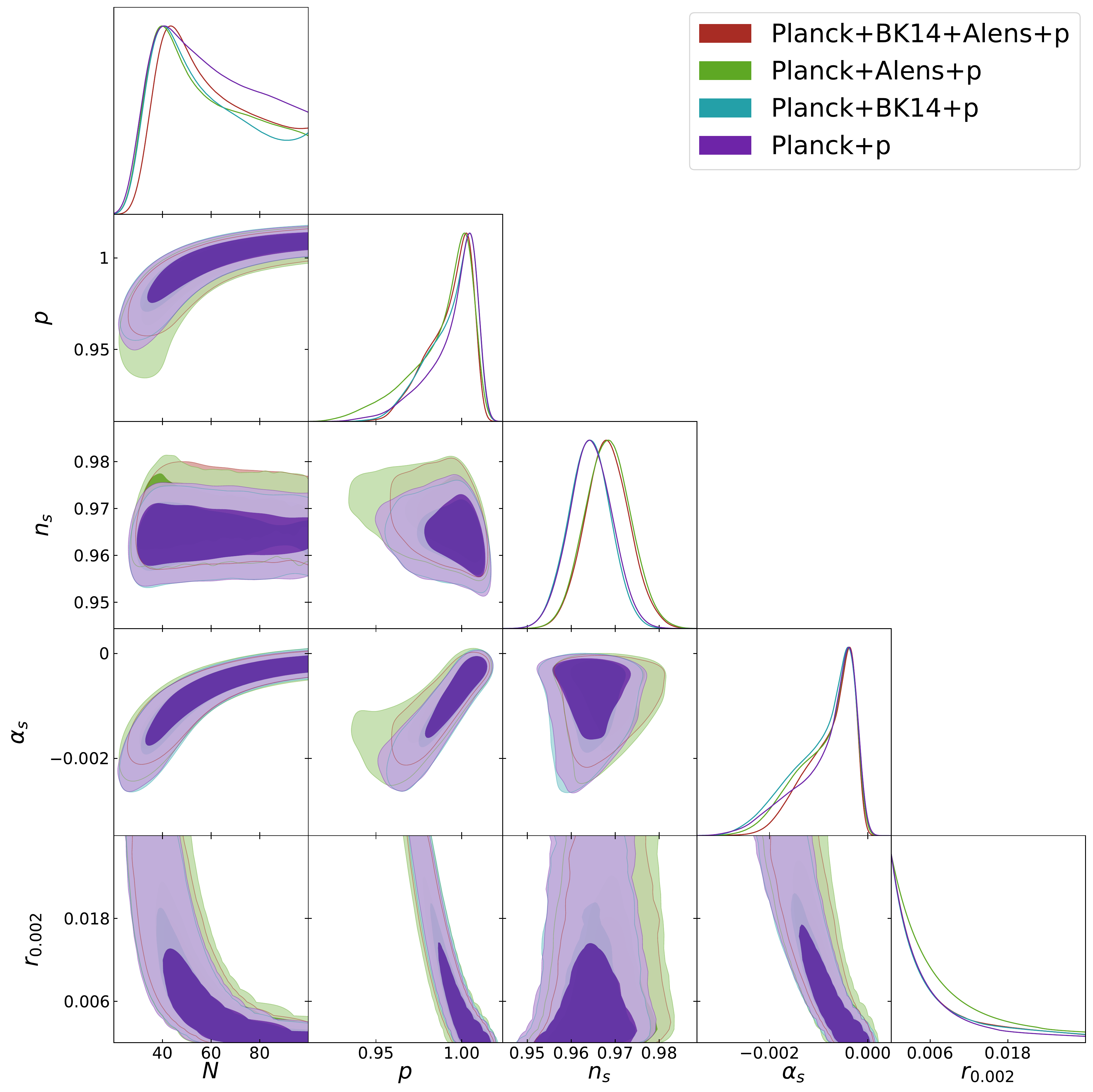}
	\caption{Constraints at $ 68\% $ and $ 95\% $ C.L. for the full Planck 2015 likelihood (Planck) and its combination with the Biceps/Keck 2015 B-mode likelihood (Planck+BK14) for the inflationary parameters for $R^{2p}$ models}
	\label{Fig.StarOb_results_with_p}
\end{figure*}

We report the constraints on the inflationary parameters for general $R^{2p}$ model with $p\simeq 1$ in Table \ref{Tab.StarOb_results_with_p}. The $ 68\% $ and $ 95\% $ C.L. contour plots are showed in Fig.\ref{Fig.StarOb_results_with_p} instead. We start again discussing the results from the Planck datasets alone (without the inclusion of $A_{\rm lens}$) reported in the first column of Table \ref{Tab.StarOb_results_with_p}. As expected the inclusion of the index $p$ in the analysis does not significantly modify the bounds on the standard cosmological parameters ($\Omega_b h^2$, $\Omega_c h^2$, $A_\mathrm{S}$, $n_S$ and $\tau$) coming from the Planck datasets alone. Conversely the constraints on inflationary parameters are largely changed by the inclusion of the index $p$. When $p$ is varied,  the number of e-folds of inflation are basically unconstrained within the flat range we imposed in our runs while the 2-$\sigma$ bound on the tensor-to-scalar ratio is relaxed to only an upper bound. We note however that the upper limit on $r$ is halved with respect to the bound reported in the Planck 2015 release ($r < 0.11$), again this is due to the correlation between the inflationary parameters arising in $R^{2p}$ inflationary models. The bound on $\alpha_S$ is also worsen by a factor $\sim 4 $ leading to a running consistent with zero nearly at 2-$\sigma$ level. Interestingly instead we are able to constraints the index $p$ with an accuracy of the $ 0.2\% - 0.3\% $. In order to understand why this is happening we should look again at Fig.\ref{fig2}. As we can see from the left panel of Fig.\ref{fig2}, for arbitrary small value of $r$, the scalar index saturates to a constant value which is only a function of the index $p$ (see also Eq.\eqref{Eq.ns_r}). For $1.0 \lesssim p \lesssim 1.02$, the saturation value of $ n_S $ falls well within the Planck constraints for $r\rightarrow 0$ therefore for these models we do not find any lower limit on the amplitude of tensor modes. For $p > 1.02$ the value of $n_S$ is always outside the Planck bound making these models incompatible with Planck data, instead models with $p < 1.0 $ are compatible with Planck data only for value of the tensor-to-scalar ratio in the range $10^{-2} < r < 10^{-1}$. This behavior of the scalar index for different value of $p$ leads to the highly non-Gaussian posteriors for $p$ and $\alpha_s$ of Fig.\ref{Fig.StarOb_results_with_p} and to the disappearance of the lower bound on $r$. Including $\alens$, we again see the shift in the best fit values of $A_S$ and $n_S$ as for the case where $p$ is kept fixed leading to a worsening of the limit on $r$ of the $45\%$ and of the constraints on $p$ of the $20\%$.
We see from the third column of Table \ref{Tab.StarOb_results_with_p} and Fig.\ref{Fig.StarOb_results_with_p} that the combination of Planck and BK14 datasets improves sligtly the upper limit on tensor amplitudes while the other parameter bounds are virtually unchanged. The inclusion of $\alens$ now only changes the bound on $\alpha_S$ shifting the best-fit toward zero by the $14\%$ and improving the 2-$\sigma$ constraints by the same amount. Again we notice that the inclusion of $\alens$ provides a better fit to the data with $\Delta\chi^2 \simeq 4-5$ 
\begin{table*}[!hbtp]
	\centering
	\begin{tabular}{lcccc}
		\toprule
		\horsp
		    &                                   Planck+p                                      \vertsp Planck+p+$A_{\rm lens}$              \vertsp Planck+BK14+p                    \vertsp Planck+BK14+p+$A_{\rm lens}$ \\
		\hline\hline
		\morehorsp
		$\Omega_{\mathrm{b}} h^2$               \vertsp $ 0.02223\pm 0.00016 $ 				  \vertsp $ 0.02240\pm 0.00018 $               \vertsp $ 0.02223\pm 0.00016 $             \vertsp $ 0.02240\pm 0.00018$         \\  
		$\Omega_{\mathrm{c}} h^2$               \vertsp $ 0.1198\pm 0.0015 $ 				  \vertsp $ 0.1184\pm 0.0016 $                 \vertsp $ 0.1201\pm 0.0015 $               \vertsp $ 0.1184\pm 0.0016 $         \\  
		$\ln(10^{10} A_\mathrm{S}) $            \vertsp $ 3.092\pm 0.033 $ 				      \vertsp $ 3.043\pm 0.041 $                   \vertsp $ 3.101\pm 0.032 $                 \vertsp $ 3.046\pm 0.041 $         \\  
		$ N $  			                        \vertsp $ 60^{+40}_{-30}$ 				      \vertsp $ 59^{+40}_{-30} $                   \vertsp $ 59^{+40}_{-30} $                 \vertsp $ 61^{+40}_{-30} $         \\  
		$ p  $  		                        \vertsp $ 0.995^{+0.021}_{-0.033} $ 	      \vertsp $ 0.990^{+0.025}_{-0.039} $         \vertsp $ 0.994^{+0.021}_{-0.030} $         \vertsp $ 0.993^{+0.020}_{-0.027} $       \\
		$ n_S $  		                        \vertsp $ 0.9644\pm 0.0049 $ 			      \vertsp $ 0.9683\pm 0.0051 $                 \vertsp $ 0.9640\pm 0.0049$                \vertsp $  0.9683 \pm 0.0051 $       \\
		$ \alpha_S $                             \vertsp $ -0.00084^{+0.00084}_{-0.0013} $    \vertsp $ -0.00083^{+0.00080}_{-0.0011} $   \vertsp $ -0.00090^{+0.00088}_{-0.0013}$  \vertsp $ -0.00077^{+0.00072}_{-0.0010} $\\
		$ r_{0.002} $                           \vertsp $  < 0.0515 $	                      \vertsp $ < 0.0750 $                         \vertsp $ < 0.0483 $                       \vertsp $ < 0.0422 $  \\
		$ \tau $  		                        \vertsp $ 0.079\pm 0.017 $ 			          \vertsp $0.056\pm 0.020 $                    \vertsp $ 0.082\pm 0.017 $                 \vertsp $ 0.057\pm 0.020 $       \\	
		$ \chi^2 $                              \vertsp $ 12949 $ 						      \vertsp $ 12945 $                            \vertsp $ 13595 $                          \vertsp $ 13590 $     \\
		\bottomrule
	\end{tabular}
    \caption{Constraints on inflationary parameters for near-Starobinsky inflation ($ p \simeq 1 $) from the Planck and Planck+BK14 datasets with and without the inclusion of the parameter $ A_{\rm lens} $. Constraints on parameters are at the $ 68\% $ C.L. for $\Omega_{\mathrm{b}} h^2, \Omega_{\mathrm{c}} h^2$ and $A_S$ while constraints on $\alpha_S, N$ and $p$ are at $95\%$ C.L. since their posteriors are highly non-Gaussian. Upper bound are also at $95\%$ C.L.}
    \label{Tab.StarOb_results_with_p}
\end{table*}

\section{Conclusions}
In this paper, we have obtained constraints on inflationary parameters using a set of recent CMB data and under the assumption of the Starobinsky model. We have also considered a particular class of inflationary models that generalize Starobinsky inflation and the possibility of an extension to $\lcdm$ described by the $A_{lens}$ parameter. 

We can summarize our results as follows : 

\begin{itemize}

    \item When conservatively considering Starobinsky inflation, corresponding to $p = 1$, and using the full Planck 2015 likelihood we obtain an upper limit on the tensor to scalar ratio $r > 0.0017$ at $95\%$ C.L. and an indication for a negative running at more than two standard deviations.  While smaller values for $r$ are allowed, also values of $r\sim 0.006$ are now inside the
    $95 \%$ C.L. Interestingly, models with a larger value of $r$ would also predict a more negative value of the running $\alpha_s$.
    The maximum value of $\alpha_s\sim-0.001$ (see Figure~\ref{Fig.StarOb_results}), however, is not within the reach of the future CMB-S4 experiment that is expected to have a sensitivity on the running of $\Delta \alpha_s \sim 0.0026$~\cite{CMB-S4}. The combination of the Planck and BK14 datasets leaves our results almost unchanged. As discussed above, this is related to the fact that our results are coming from the Planck bound on $n_S$ and from assuming inflationary consistency relations between $n_S$, $r$ and $\alpha_S$ and therefore they are not significantly affected from the inclusion of the BICEP2 B-mode likelihood.
    
    \item Considering the phenomenological lensing parameter $\alens$ shifts the best-fit values of $r$ and $\alpha_S$ due to the degeneracy between $\alens$ and the scalar parameters $n_S$ and $A_S$. When $\alens$ is considered, the upper limit is now $r>0.0013$ at $95 \%$ C.L., i.e., the amount of gravitational waves predicted is significantly smaller. Future CMB experiments should, therefore, target to a $\Delta r \sim 0.0003$ sensitivity if they plan to falsify the Starobinsky model at the level of five standard deviations. This sensitivity is about a factor two better than the one predicted for the CMB-S4 experiment. 
    
    \item For a more general $R^{2p}$ inflation and using the full Planck likelihood, we found no lower limit for the tensor mode amplitude. Conversely, we obtain a tight constraint on the index $p$ at the $95\%$ C.L. confirming that small departures from the Starobinsky model are allowed by the Planck data with values in the range $0.962 \leq p \leq 1.016$. The inclusion of $\alens$ worsen this constraint by the $20\% $. When considering the combination of the full Planck dataset with the BK14 dataset again we do not find any improvement w.r.t. to the Planck datasets alone. However, including $\alens$ now do not worsen the constraints on $p$ but only shift the best fit of $\alpha_S$ to a less negative value.  
    
\end{itemize}

We, therefore, confirmed that Starobinsky inflation provides an excellent fit to the most recent data, but that uncertainties on $n_s$ and on the value of $A_{lens}$ could easily bring the expected value of $r$ in the region of $r\sim0.001$. If the primordial inflationary background is at this level, it will not be detectable either by the Simons Observatory~\cite{Ade:2018sbj}, that has an expected sensitivity around $\Delta r \sim 0.002$, either by the LiteBIRD satellite that is planned to have a sensitivity of $\Delta r \sim 0.001$. It will also be barely detectable by CMB-S4~\cite{CMB-S4} that is expected to reach a target sensitivity of $\Delta r \sim 0.0006$. Moreover, the goal of the CMB-S4 mission to "achieve a $95 \%$ confidence upper limit of $r < 0.001$"~\cite{CMB-S4} can be severely affected if the primordial gravitational waves background is in the region of $r\sim0.001$. 

However, values of $r$ could also reach the $r\sim 0.006$ region, allowing, in this case, a statistically significant detection at about three standard deviations for the Simons Observatory and at about ten standard deviations for CMB-S4. In the optimistic case of $r\sim 0.006$ we also expect a running of the spectral index $\alpha_s\sim-0.001$. Unfortunately this value can't be detectable even by future CMB experiments as CMB-S4 (with expected sensitivity of $\Delta \alpha_s\sim0.002$ \cite{CMB-S4}), but it could be reachable when information from future lensing or galaxy clustering measurements are included.
Small departures from the Starobinsky model are also possible and in agreement with observations. In this case, we found no predicted lower limit to $r$.

We conclude noting that the inflationary prediction on curvature perturbation may be spoiled by a reheating phase accompanied by some parametric resonance (see e.g. \cite{Moghaddam:2014ksa,Jiang:2018uce}). This process can also take place in a Starobinsky-like inflation as showed in \cite{Fu:2019qqe}. While the prediction on the inflationary observables coming from the post-inflationary evolution of the scalar field strongly depends on the inflationary model under consideration and on the coupling between the inflaton and the entropy field responsible for the reheating mechanism, the general outcome of such a reheating phase is to suppress the value of the tensor-to-scalar ratio by enhancing the amplitude of primordial density fluctuations. The reheating phase can therefore significantly modify the prediction in the $(n_s - r)$ plane for a chosen inflationary model. However, this does not apply to the constraints drawn in this work. In the present analysis, infact, the bound on $r$ comes from the functional dependence imposed by the $R^{2p}$ models between $r$ and $n_s$. Since the value of $n_s$ is tightly constrained by the Planck data so it is the value of $r$ provided that Eqs.\eqref{eq.r_ns_alphas} hold. A successful Starobinsky-like model including a reheating phase must be able to predict an amplitude of scalar fluctuations consistent with Planck data leading to the same constraints for $n_s$ and $r$ we found in this work. Conversely the requirement that the reheating phase must give a value of $A_s$ compatible with Planck data can be used to place strong constraint on the reheating mechanisms. However a detailed study of the reheating phase in Starobinsky-like inflationary models is out of the scope of the present paper and we left if for a future work.

\acknowledgments
\noindent
We thank Gaudalupe Cañas Herrera, Kaloian Lozanov and Fabio Moretti for useful comments and discussions. AM thanks the University of Manchester and the Jodrell Bank Center for Astrophysics for hospitality. AM and FR are supported by TASP, iniziativa specifica INFN, Italy. FR also acknowledges support from the NWO and the Dutch Ministry of Education, Culture and Science (OCW), and from the D-ITP consortium, a program of the NWO that is funded by the OCW.

\bibliographystyle{aipnum4-1}
\bibliography{biblio}

\end{document}